\documentclass[conference]{IEEEtran}

\usepackage{cite}
\usepackage{threeparttable}
\usepackage{amsmath,amssymb,amsfonts}
\usepackage{algorithm, algorithmic}
\usepackage{graphicx}
\usepackage{textcomp}
\usepackage{balance}
\usepackage{xcolor}
\def\BibTeX{{\rm B\kern-.05em{\sc i\kern-.025em b}\kern-.08em
    T\kern-.1667em\lower.7ex\hbox{E}\kern-.125emX}}
\begin{document}

\title{Tablext: A Combined Neural Network And Heuristic Based Table Extractor}

 \author{\IEEEauthorblockN{Zach~Colter}
 \IEEEauthorblockA{\textit{Department of EECS} \\
 \textit{University of Michigan}\\
 Ann Arbor, US \\
 zcolter@umich.edu}
 \and
 \IEEEauthorblockN{Morteza~Fayazi}
 \IEEEauthorblockA{\textit{Department of EECS} \\
 \textit{University of Michigan}\\
 Ann Arbor, US \\
 fayazi@umich.edu}
 \and
 \IEEEauthorblockN{Zineb~Benameur-El}
 \IEEEauthorblockA{\textit{Department of EECS} \\
 \textit{University of Michigan}\\
 Ann Arbor, US \\
 zinebbe@umich.edu}
 \and
 \IEEEauthorblockN{Serafina~Kamp}
 \IEEEauthorblockA{\textit{Department of EECS} \\
 \textit{University of Michigan}\\
 Ann Arbor, US \\
 serafibk@umich.edu}
 \and
 \IEEEauthorblockN{Shuyan~Yu}
 \IEEEauthorblockA{\textit{Department of EECS} \\
 \textit{University of Michigan}\\
 Ann Arbor, US \\
 shuyan@umich.edu }
 \and
 \IEEEauthorblockN{Ronald~Dreslinski}
 \IEEEauthorblockA{\textit{Department of EECS} \\
 \textit{University of Michigan}\\
 Ann Arbor, US \\
 rdreslin@umich.edu}
 }
\maketitle

\begin{abstract}
A significant portion of the data available today is found within tables. Therefore, it is necessary to use automated table extraction to obtain thorough results when data-mining. Today's popular state-of-the-art methods for table extraction struggle to adequately extract tables with machine-readable text and structural data. To make matters worse, many tables do not have machine-readable data, such as tables saved as images, making most extraction methods completely ineffective. In order to address these issues, a novel, general format table extractor tool, Tablext, is proposed. This tool uses a combination of computer vision techniques and machine learning methods to efficiently and effectively identify and extract data from tables. Tablext begins by using a custom Convolutional Neural Network (CNN) to identify and separate all potential tables. The identification process is optimized by combining the custom CNN with the YOLO object detection network. Then, the high-level structure of each table is identified with computer vision methods. This high-level, structural meta-data is used by another CNN to identify exact cell locations. As a final step, Optical Characters Recognition (OCR) is performed on every individual cell to extract their content without needing machine-readable text. This multi-stage algorithm allows for the neural networks to focus on completing complex tasks, while letting image processing methods efficiently complete the simpler ones. This leads to the proposed approach to be general-purpose enough to handle a large batch of tables regardless of their internal encodings or their layout complexity. Additionally, it becomes accurate enough to outperform competing state-of-the-art table extractors on the ICDAR 2013 table dataset.
\end{abstract}
\maketitle

\section{Introduction}
Tables are widely used in documents, articles, web-pages, etc. as they can concisely show complex information in a way that is suitable for human readers~\cite{tengli2004learning}. Autonomous table extraction enables translating this abundant amount of information to a machine-readable format which has broad applications in data-mining and information-retrieval~\cite{oro2009trex,pinto2003table}. 
The complexity of some tables' layouts and the various formats in which they may be found are current problems that prevent the existence of a universal, highly-accurate, automatic table extraction tool.

Formats such as Hypertext Markup Language (HTML), Portable Document Format (PDF), and Portable Network Graphics (PNG) are common encodings that may contain tables. Because of the many different ways to represent a table, designing separate extraction methods for every encoding is time-consuming and prone to errors. Moreover, an environment may have multiple formats that are not encoded uniformly. For instance, a web-page may contain an image that is not encoded in HTML. Unifying internal encodings of files is the first step to overcome universality issues.

Tablext solves the universality issue by extracting data from images. There are several Java and Python open-source libraries to convert specific file formats to images such as pdf2image~\cite{pdf2image} (PDF to PNG), GrabzIt~\cite{GrabzIt} (HTML to PNG), etc. If a tool does not exist to convert an obscure format, a user can always take screenshots of the desired file.

Tablext uses separate Convolutional Neural Networks (CNNs)~\cite{lecun1999object,fayazi2021applications} to identify and extract information from tables, but it does not solely rely on them. Before data extraction, Tablext uses common conventional computer vision techniques, such as line identification, to extract positional information from tables before another CNN is used to fix small mistakes. This approach can be both faster and more accurate than having a single neural network. 

By first identifying important features within the tables, each neural network only has to solve a small sub-task. This means that the size required for each network is drastically reduced. The smaller size of the neural networks reduces the computation requirement and also reduces the risk of overfitting.

Tablext begins by using an efficient CNN to identify the exact location of the tables within the input images. Using an ensemble of this CNN and a object detection network, YOLO \cite{yolo}, allows for the extraction algorithm to be confined to relevant areas. By isolating the relatively expensive extraction algorithm to areas found by the identification CNN, Tablext is efficient in handling massive databases.

Tablext achieves a high extraction accuracy by both identifying the lines separating various cells, as well as looking at the positions of the text within the tables. Together, these methods can recognize complex table structures. It can generally be assumed that the identified lines will correctly split the table into cells, but when no lines exists or they cannot be correctly identified, the positional data is used to separate the cells.  After correctly segmenting the table into cells, the data is cleaned by a CNN before the open-source Optical Character Recognition (OCR) software, Tesseract~\cite{tesseract}, identifies the text in every individual cell.

The novel contributions of this paper can be summarized as follows.
\begin{itemize}
  \item An new method to identify cells that utilizes the positions of the data and lines in order to obtain high precision preliminary results at a low
  computational cost. 
  \item A hybrid pipeline that allows the neural network to focus its attention on complex problems, resulting in increased accuracy.
  \item The novel code will be released open-source after the publication.
\end{itemize}

The rest of the paper is organized as follows. Section~\ref{sec:Related_Work} briefly presents related works. Section~\ref{sec:Table_Identification} describes the proposed method for identifying the location of the tables. Section~\ref{sec:Table_Extraction} explains Tablext's approaches for data extraction. Section~\ref{sec:Results} shows the final results and comparisons to other table extractors. Finally, the paper is concluded in Section~\ref{sec:Conclusion}. Throughout this paper, four different images with tables are used to explain various concepts. For clarity, figures will declare which original images they are referencing.

\section{Related Work}
\label{sec:Related_Work}
Several methods have been studied table identification and extraction. However, some of them describe various approaches to identify tables from different types of documents and do not focus on data extraction \cite{Ident0, Ident1, Ident2, Ident3, Ident4, Ident5, Ident6}. 

\cite{perez2016tao,Hassan2007,Yildiz2005pdf2tableAM,oro2009trex,tableseer} propose systems for the identification and extraction of tables within PDF files. Oro and Ruffolo~\cite{oro2009trex} find positional data for the text and the use an agglomerative hierarchical clustering algorithm~\cite{jain1999data} to combine words into lines. The main limitation of this approach is that it only supports unlocked, machine-readable, PDFs and a large amount of tables are found in locked PDFs or images. Perez-Arriaga et al.~\cite{perez2016tao} find columns and rows by comparing the locations of the text-boxes and combining text-boxes below a distance threshold. Although the method has a decent recall, its false positive rate is high. Liu et al.~\cite{tableseer} propose a method in which, tables are found by grouping text with similar positions and font size. This approach works with HTML format in addition to PDF files. However, it is only optimized for research papers and its extraction performance is poor especially on general documents. Similar to the other mentioned studies, this method cannot work with locked PDF files or images. 

The following papers focus on table extraction that are not related to PDFs. Tengli et al.~\cite{tengli2004learning} propose a ML-based technique for table extraction from HTML files. Nishida et al.~\cite{html2017} focuses on HTML extraction too. They use a Recurrent Nerual Network (RNN) model to extract data from HTML tags. These such approaches are not applicable to formats other than HTML and are therefore, limited in scope. Pinto et al.~\cite{pinto2003table} propose a method to identify the horizontal lines within tables, however they do not concentrate on table extraction.

Koci et al. and Puha et al.~\cite{8395185,Puha2018EnhancingOD} propose methods that can handle tables in more diverse formats. Koci et al.~\cite{8395185} propose a graph representation of tables so that they can use heuristics approaches in order to extract information. Their proposed methodology is optimized to extract data from various spreadsheets. Puha et al.~\cite{Puha2018EnhancingOD} propose a heuristic method for general table extraction. They recognize tables in images by first using OCR to identify data within cells, then using the locations of the data.

Many state-of-the-art general table extraction methods solely use neural networks. Qasim et al.~\cite{graph_nn} use a graph neural network to identify cell locations within tables. DeepDeSRT~\cite{dualNetworks}, uses separate, specially made neural networks to identify tables and extract their data. While TableNet~\cite{SingleNetwork}, uses a single network for both identification and extraction. They argue this is an efficient approach since these tasks are interdependent. However, this also means that the network is less optimized for both tasks. Tablext outperformed both of these methods when tested on a common dataset. 

By utilizing machine learning, modern table extractors can handle more diverse tables than earlier works. However, identifying and extracting data is a difficult undertaking that requires many sub-tasks to be completed. Many of these tasks, such as identifying lines that can make up a table's structure, are good candidates for conventional computer vision methods. Instead of forcing a neural network to learn and solve every one of these tasks, our paper introduces a new approach that combines both conventional computer vision and machine learning techniques. This allows the neural network to be smaller in size and focused on its particular task of identifying abnormal cells. This is especially important given the limited amount of labeled training data available for tables.
\section{Table Location Identification}
\label{sec:Table_Identification}
While the main focus of this paper is table extraction, a novel algorithm was created and used to identify tables before extraction. It is imperative to find the table locations before extraction because doing so increases both the performance and accuracy of the algorithm. The boost in performance comes from reduced search space for the complex information extraction. The increased accuracy is a result of the table identification network separating tables from one another and removing irrelevant text and lines that could potentially confuse the table extractor. To further increase the accuracy of the table identification, it is later combined with YOLO.

The identification algorithm needs to find both the minimum and maximum $X$ and $Y$ coordinates for each table that exists within each page. To efficiently calculate these coordinates, two relatively simple CNNs are used in succession. These two networks identify at which $Y$ and $X$ coordinates, respectively, a table exists.

All input images, after being converted from any file format, are first converted to gray-scale and resized. Preliminary tests show that a width of $800$ pixels allows for accurate table extraction without a significant performance penalty. The height at this stage is scaled proportionally to avoid distortion. This resized image cannot directly be sent to a conventional CNN because its height is variable. Additionally, the large size of the input data would require a significant amount of memory for each batch, slowing down the training process. In order to reduce the input size and make the input regular, the image is sliced into horizontal strips, each with a height of $64$ pixels. This number is chosen to reduce the training time without significantly degrading accuracy.

While cutting the image into strips reduces the input and output size of the network, this process introduces a new problem. The accuracy of identifying a table at the top and the bottom of the slices is lower than identifying a table near the middle. Intuitively this is because the network only has information in one direction from these edge outputs. To combat this issue, each strip only detects the existence of a table within its innermost $32$ rows of pixels. To cover the entire table, a moving window approach is used where each input strip partially overlaps its adjacent strips.

The neural network has $4$ output nodes for every horizontal strip. Each node determines if there is a table at a particular location within the centermost $32$ pixels. In the innermost $32$ pixels, node $i, 0 \leq i \leq 3,$ of the neural network is trained to identify if a table exists between any of the $[8*i, 8*i+7]$ pixels. The outputs of all the slices are concatenated and stored in an array that identifies in which rows a table exists. With this approach, the top and bottom of each page cannot be searched for tables, because only the inner pixels are represented in the outputs. To fix this problem, the input image is extended by $16$ pixels at the top and bottom with a white border.

\begin{figure}
\begin{center}
\includegraphics[width=\columnwidth]{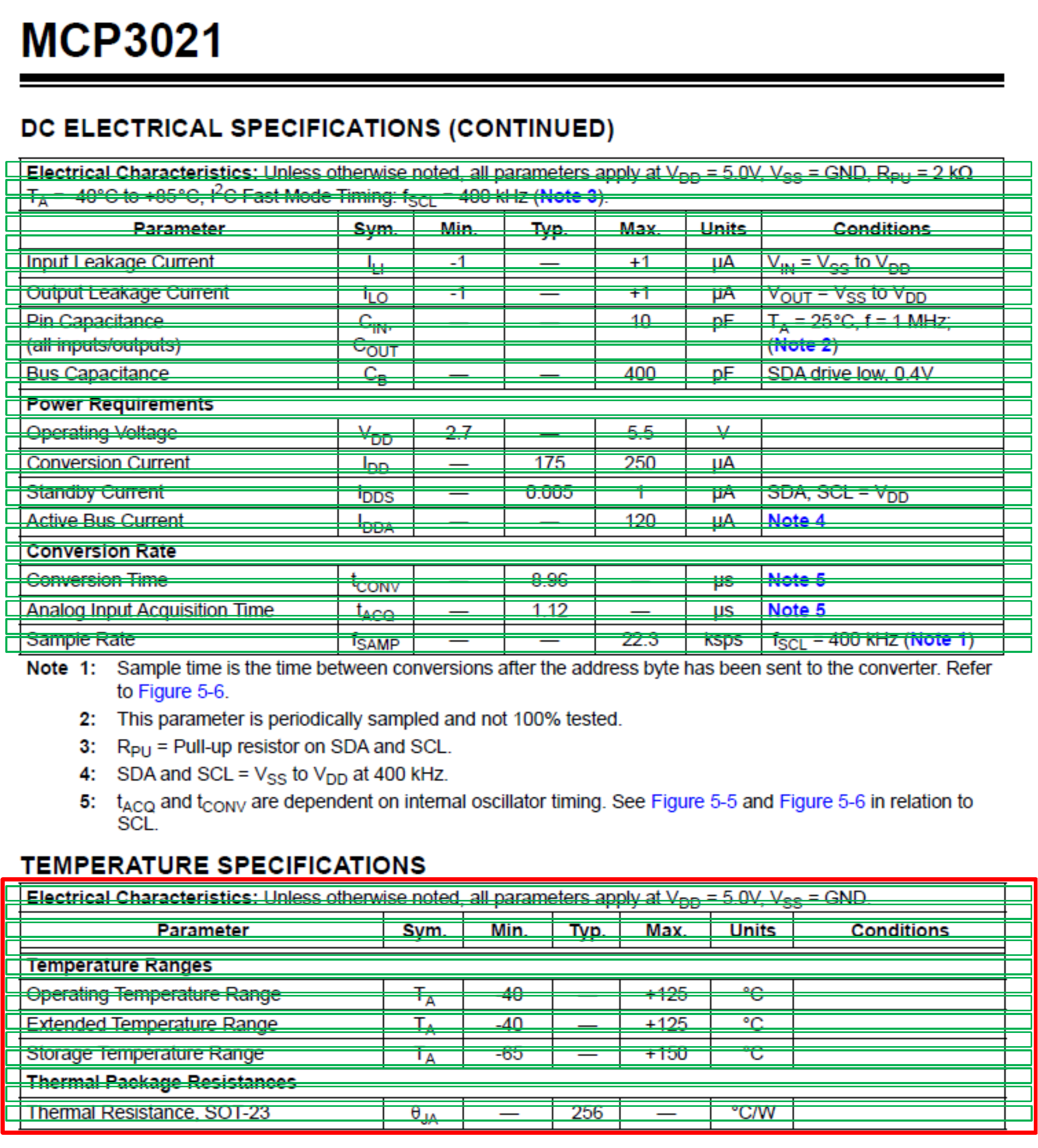}
\caption{(Table A) Horizontal boxes each representing a row where a table is identified. The bottom group is highlighted with a thick box (red) for clarification.}
\label{fig:post_ver}
\end{center}
\end{figure}

Figure~\ref{fig:post_ver} shows the concatenated output of all the slices from the neural network. In this figure, the thin horizontal boxes represent every row the network checked for a table. These boxes show the $Y$ locations where the neural network predicts a table exists. While this particular example shows a table that takes up the entire width of the document, this network can also identify tables that are within one of several columns on the page.

The next step is to clean the output by concatenating rows of positively identified tables into groups. If no groups are created, because none of the rows contain a table, Tablext outputs that no table is found in the image. Otherwise, the groups are prepared and sent to the second neural network in order to find the correct $X$ coordinates. This process is simpler than having to find the correct $Y$ coordinates as the amount of area to search is smaller. For this reason, a CNN is used without a sliding window approach.

\begin{figure}
\begin{center}
\includegraphics[width=\columnwidth]{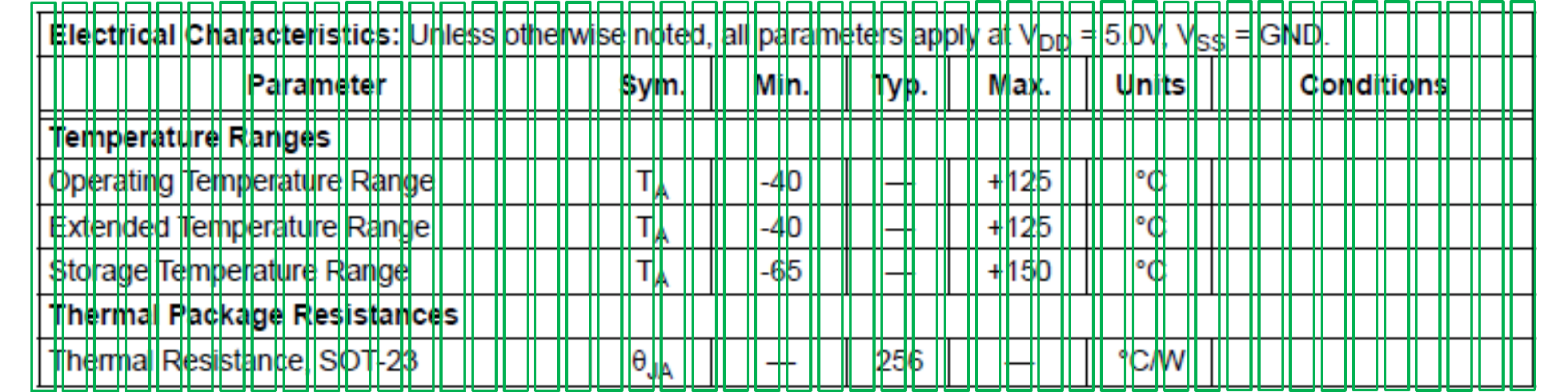}
\caption{(Table A) Vertical boxes representing every column where a table is identified. }
\label{fig:post_hor}
\end{center}
\end{figure}

Each group of rows is resized into a $400 \times 400$ pixel square to efficiently achieve high accuracy. This input size is obtained by preliminary testing. The image is sent into a CNN that identifies which columns contain a table. Figure~\ref{fig:post_hor} shows this technique being applied to the bottom group of rows in Figure~\ref{fig:post_ver}. The actual result produced by the network looks distorted to a human because the table is scaled into a square. Instead a representative, recreated image is shown. Here the vertical boxes show where the network predicts that a table exists. If the table is only in a single column of a multi-column image, the lines would cover only that section. This network can also identify two separate adjacent tables. This is achieved by not identifying a table in between the two table groups. Each vertical group is separately sent to this second stage, as an example Figure~\ref{fig:post_ver} would send three separate images into the second network.

\begin{figure}
\begin{center}
\includegraphics[width=\columnwidth]{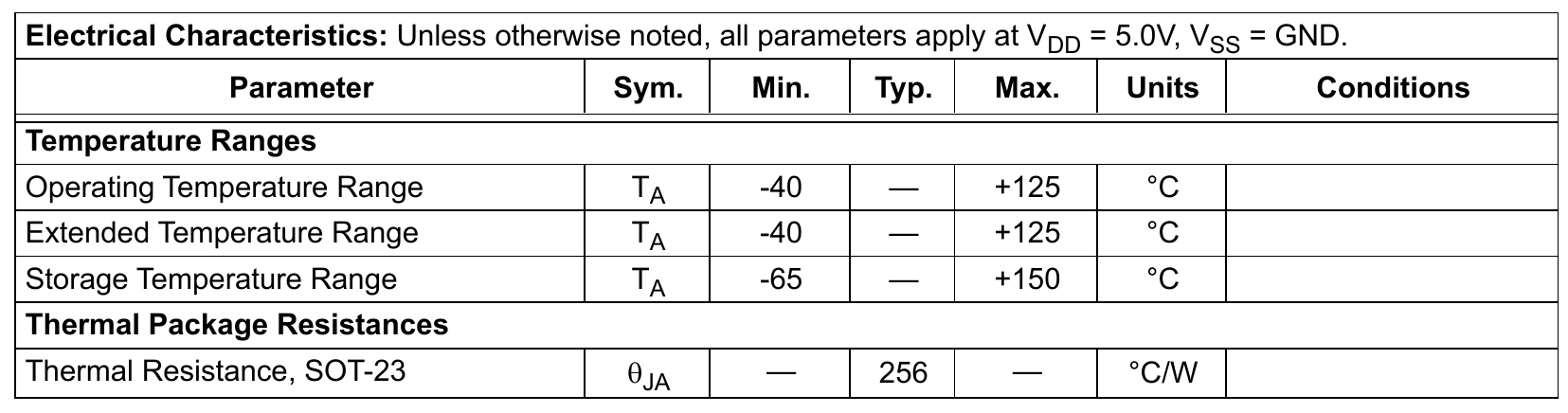}
\caption{(Table A) A cutout of the original table using the coordinates identified.}
\label{fig:final_ident}
\end{center}
\end{figure}

The columns are concatenated into groups, in a similar manner to the rows. The scaling that converted the group of rows into the square, shown in Figure~\ref{fig:post_hor}, is inverted and applied to the $X$ coordinates. Additionally, the scaling used to convert the original image into the resized image, shown in Figure~\ref{fig:post_ver}, is inverted and applied to the $Y$ coordinates. Doing this gives accurate coordinates of each table's start and end position regarding the original image. Figure~\ref{fig:final_ident} shows a cutout of the first table, using the found $X$ and $Y$ coordinates.

The CNN by itself tends to over propose table regions because the existence of lines on the page are very influential to the model. This causes the CNN to predict areas around these lines as tables whether these lines correspond to tables or not. This will cause the extraction portion of the algorithm to take more time as it has to run OCR on these extra regions to figure out they are not tables. To counteract this, a YOLO \cite{yolo} network is trained on table images to see if this more popular object detection method could outperform the CNN. YOLO typically identifies all the right table regions, but the bounding boxes that it proposes do not cover the entire table. For this reason, an ensemble of the CNN and YOLO is used to create an optimal identification model. Both networks first independently propose table regions. When these models propose regions that overlap, the union of these two regions is taken. This ensures both that the final regions are likely to contain tables and that the regions will likely contain the entirety of the tables.

While this approach can perfectly identify tables in multiple columns, there is an extremely rare case where two adjacent tables in separate columns have vastly different heights. In this case, extra data could be identified as a table. This rare issue is rectified later because the proposed extraction method can tolerate some extra area identified as a table.

\section{Table Extraction}
\label{sec:Table_Extraction}
After all of the tables are identified in the input file, the extraction method starts independently on each table. In order to process high resolution images efficiently, a resized copy of each individual table is created with a width of $800$ pixels and a proportional height. This copy is used for a majority of the cell identification steps, however the original image is kept because it is used later during OCR.

\begin{figure}
\begin{center}
\includegraphics[width=.4\textwidth]{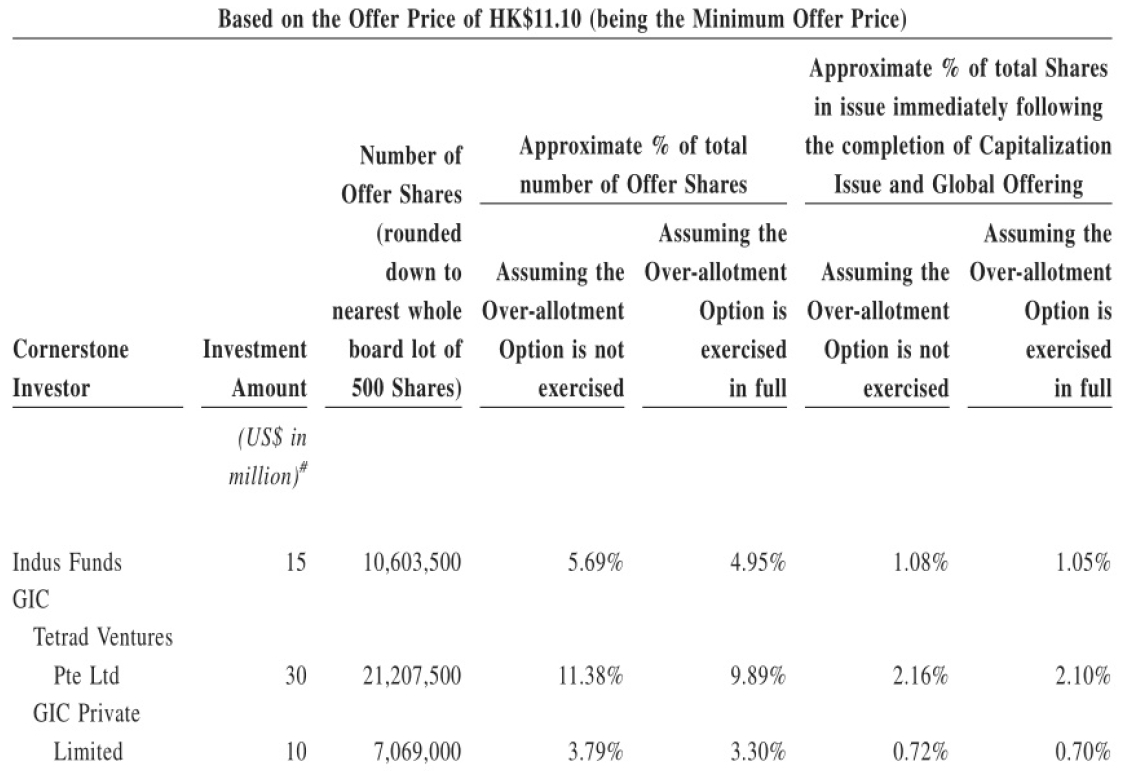}
\caption{(Table B) A table sample identified by the proposed table identification method.}
\label{fig:extract_original}
\end{center}
\end{figure}

Several methods are used to identify and isolate the cells within each table. The table extraction begins by identifying the high-level, general structure of the table. It identifies the vertical and horizontal lines, both visible and implied, that correctly divide a majority of the cells. There are several ways to locate this information within the tables. One approach is to identify the visible lines separating various cells. Another method is to look at the positions of the text within the tables and extrapolate how the cells should be positioned in accordance to each other. Tablext uses both of these approaches in order to obtain the structure of the table. Afterwards, a neural network is used to clean up any cells that do not conform to the general structure. Figure~\ref{fig:extract_original} shows a table that already went through table identification. This will be used as an example to explain the steps of the table extraction. 

\subsection{Real Line Identification}
Vertical and horizontal lines are commonly used to separate and organize the contents of tables. To identify these lines, the second-derivative gradient field is approximated for the gray-scale image of the table. Assuming that $f(x,y)$ represents the brightness of a pixel at the location $(x,y)$. The values of $\frac{\partial^2}{\partial x^2}f(x,y)$ and $\frac{\partial^2}{\partial y^2}f(x,y)$ are approximated for each pixel by looking at their adjacent pixels and the gradients of those pixels. These gradients are then stored into separate matrices. Then a $3\times3$ maxpooling is independently applied to these matrices, to account for noise in the image and imperfect lines. The maxpooling has a stride of $1$ in both $X$ and $Y$ directions to retain as much precision as possible when identifying the lines.

With this information, the vertical lines are found by searching for vertical segments where all the pixels in the segment have high $\frac{\partial^2}{\partial x^2}f(x,y)$ values relative to the pixels in rest of the image and similar $\frac{\partial^2}{\partial y^2}f(x,y)$ values to one another. Similarly, horizontal lines are found by searching for horizontal segments where all the pixels in the segment have high $\frac{\partial^2}{\partial y^2}f(x,y)$ values relative to the pixels in rest of the image and similar $\frac{\partial^2}{\partial x^2}f(x,y)$ values to one another.

\begin{figure}
\begin{center}
\includegraphics[width=.48\textwidth]{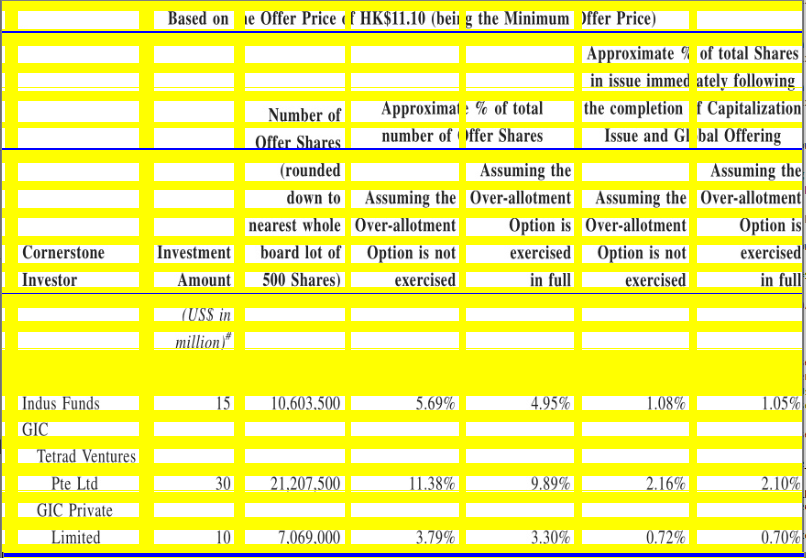}
\caption{(Table B) Inferred lines shown with bright lines (yellow) and real lines shown with dark lines (blue).}
\label{fig:extract_lines}
\end{center}
\end{figure}

The segment length has to be a small fraction of the table's height or width for the vertical and horizontal lines respectively. However, once a line is found, that line is extended to cover the entire width or height of the table. Identifying lines in this way helps define the structure of the table, but some cells that do not conform to the general format might be split by lines that do not cover the entire width or height of the table. This issue is solved later with a neural network that stitches together improperly cut cells. Figure~\ref{fig:extract_lines} is a debug image, created by the algorithm, that shows the few real lines in dark blue. The rest of the lines in this figure are created with another process that will be explained in the following section.

\subsection{Inferred Line Identification}
Real line identification alone is sufficient for regular tables that have all of their data properly split into cells with defined borders. However, as Figure~\ref{fig:extract_lines} shows, many tables only have proper borders for some of the cells or do not have any well defined lines at all. To correctly identify cells that real line identification alone could not, Tablext uses the positions of the data. Finding the locations of the data is trivial. All of the high contrast pixels, not already identified as a real line, belong to the data. With the locations of the data known, Tablext attempts to segment the pieces of data into columns and rows. To achieve this, Tablext creates straight lines that do not intersect the data. These such lines will be referred to ``inferred lines" as the table requires these non-existent lines to properly convey its information.

The task of finding inferred lines is made harder by the existence of cells that do not conform to the general table's structure. In Figure~\ref{fig:extract_original}, multiple cells span two columns. If inferred lines are defined as lines that do not overlap any pieces of data, the columns underneath the cells that span multiple columns would not be separated. In order to solve this issue, Tablext starts by defining the term ``threshold distance". 

\begin{figure}
\begin{center}
\includegraphics[width=\columnwidth]{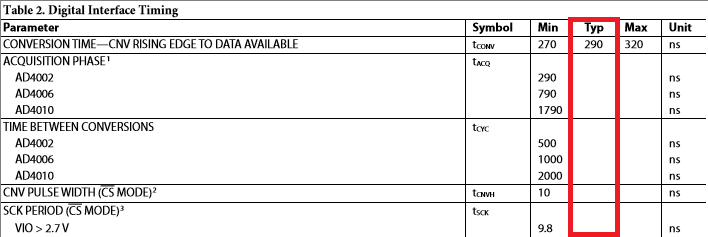}
\caption{(Table C) Initial sparse table before inferred vertical lines. A box has been placed around the, difficult to handle, referenced sparse column for clarity.}
\label{fig:Sparse_test0}
\end{center}
\end{figure}

\begin{figure}
\begin{center}
\includegraphics[width=\columnwidth]{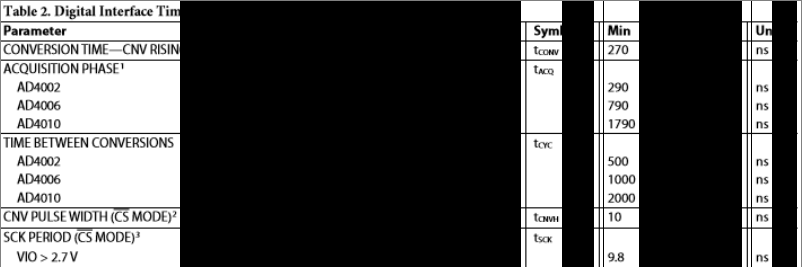}
\caption{(Table C) Sparse table with initialized inferred vertical lines.}
\label{fig:Sparse_test1}
\end{center}
\end{figure}

For vertical lines, the threshold distance is the proportion of the table's height an inferred line must not touch any data in order to be considered valid. If the table is dense, meaning that most cells contain data, a small, static threshold distance would produce accurate results. However if the table is sparse, meaning that many cells do not contain data, this approach would not work. A small, static threshold distance would combine adjacent empty cells, into wide inferred lines. Figure~\ref{fig:Sparse_test0} and \ref{fig:Sparse_test1} show this. This means that the few pieces of data within these columns or rows could be overwritten by the inferred line.

To accommodate both sparse and dense tables, the threshold distance is adaptable. Algorithm~\ref{algor:algorithm1} describes a simplified procedure of acquiring the final inferred lines with a variable threshold distance. In the algorithm, \textsf{possible inferred lines} refers to inferred lines found at the given threshold distance, \textsf{$\Delta$} is a small learned constant, and \textsf{number of groups} refers to the amount of separable groups of inferred lines that exist within the possible inferred lines.

\begin{algorithm}[h!]
\caption{\hspace{-0.1cm}\textbf{:} Setting the threshold distance}
\begin{algorithmic}[1]
    \label{algor:algorithm1}
    \STATE \textbf{Variables:} \textsf{num\_group1, num\_group2, threshold\_distance, temp\_lines}
    \STATE \textbf{Output:} \textsf{final\_lines}
    \STATE {\textbf{Start: }} 
    \STATE ~~~~~~~~~{\textsf{threshold distance = $0.6$; }\textsf{num\_group1 = 0; }\textsf{final\_lines = [];}}
    \STATE ~~~~{{\textbf{Loop: }} 
        \STATE ~~~~~~~~~{\textsf{temp\_lines = possible inferred lines;}}
        \STATE ~~~~~~~~~{\textsf{num\_group2 = number of groups;}}
        \STATE ~~~~~~~~~{\textbf{if}}  \textsf{num\_group2 $\geq$ num\_group1:}
            \STATE ~~~~~~~~~~~~{\textsf{num\_group1 = num\_group2;}}
            \STATE ~~~~~~~~~~~~{\textsf{final\_lines = temp\_lines;}}
            \STATE ~~~~~~~~~~~~{\textsf{threshold\_distance += $\Delta$;}}
            \STATE ~~~~~~~~~~~~{\textsf{go to Loop;}}

        \STATE ~~~~~~~~{\textbf{return}} final\_lines;}
\end{algorithmic}
\end{algorithm}

\begin{figure}
\begin{center}
\includegraphics[width=\columnwidth]{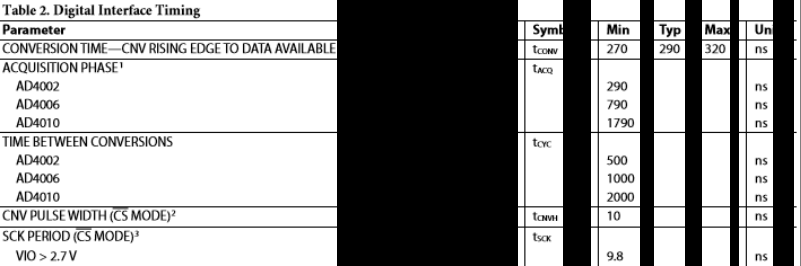}
\caption{(Table C) Sparse table with incremented inferred vertical lines.}
\label{fig:Sparse_test2}
\end{center}
\end{figure}

For a visual representation of this algorithm, images of this process are produced by the code. Figure~\ref{fig:Sparse_test2} shows the final output after applying this algorithm to the sparse table in Figure~\ref{fig:Sparse_test1}. At the start, the vertical inferred lines, represented as black columns, completely cover the column beginning with ``Typ". If the threshold distance was a low static number, this column would be merged with one of its neighboring columns. However, by using Algorithm~\ref{algor:algorithm1}, the lines, over several iterations, split around the ``Typ" column as shown by Figure~\ref{fig:Sparse_test2}.

During this process, every inferred line is assigned a ``quality score". The quality score is the percentage of the maximum possible line length that a particular inferred line covers without intersecting a piece of data. These quality scores are used later for combining the various lines.

Inferred horizontal lines are created in a similar manner as inferred vertical lines, except the threshold distance is a percentage of the table's width instead of the table's height. To better show this complete process, both vertical and horizontal inferred line identification are used on Figure~\ref{fig:extract_original}. The resulting inferred lines are used in conjunction with previously calculated real lines to create the debug image shown in Figure~\ref{fig:extract_lines}. 

\subsection{Final Structural Line Identification}
Now that all of the real and inferred lines have been located, Tablext begins to group all of this information so that it has a high level understanding of the particular table's layout. Inferred lines are combined with their neighbors, if they are within close proximity, to create different groups of inferred lines. If a real line is within a group of inferred lines, or it is only several pixels away from a group, only the real line is kept. The real lines are considered to be the ground truth, so nearby inferred line groups are redundant. If there are no real lines near an inferred line group, the quality scores of the inferred lines are used to determine which line in the group is the best line to split the cells. A Simple Moving Average (SMA) of the quality scores, looking both forwards and backwards two pixels, is calculated for each group. Lines that do not have a valid inferred line are given a zero quality score. Using an SMA helps reduce the impact of noise in the image, while also fixing corner cases where a single pixel thick inferred line might sneak between characters in a piece of data.

The index that contains the maximum SMA value is taken for each inferred group that is not near a real line. Then, the locations of the real lines and the maximum SMA indexes are concatenated together. These combined lines are the final lines for the high level table identification. This process is completed independently for both the vertical and horizontal lines. Figure~\ref{fig:extract_final} is created by applying this process on the lines found in Figure~\ref{fig:extract_lines}.

\subsection{Neural Net Correction}
\begin{figure}
\begin{center}
\includegraphics[width=.48\textwidth]{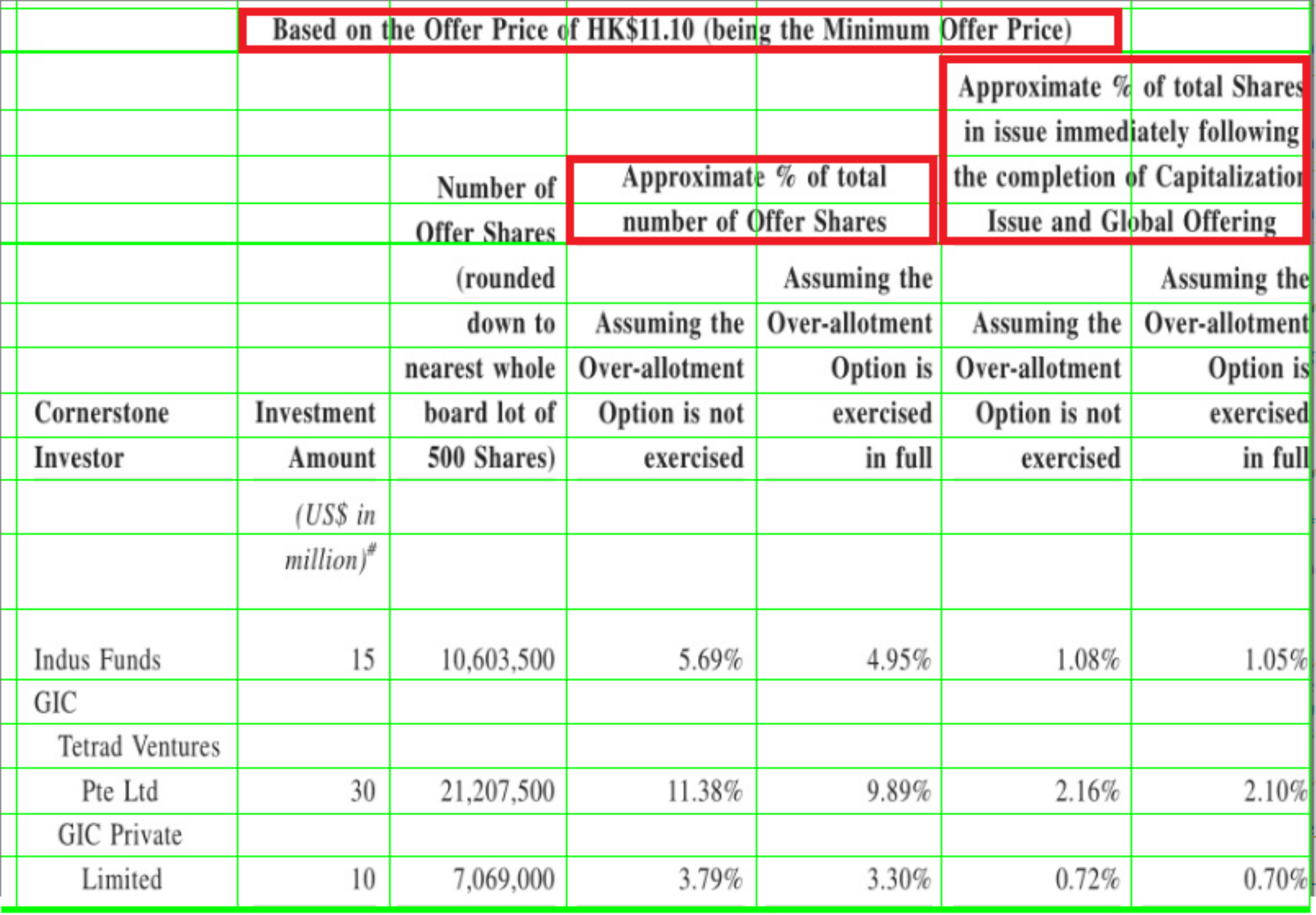}
\caption{(Table B) The merged lines (green) created by the final line identification. Vertically split cells highlighted with thick boxes (red).}
\label{fig:extract_final}
\end{center}
\end{figure}

Since the final lines describe the regular structure of the table, cells that do not conform to that structure might be sliced into two or more pieces. Figure~\ref{fig:extract_final} shows several cells that are cut in half by the final lines. In order to fix this issue, a CNN is used to recombine improperly sliced cells. A simpler to implement solution would be to check if any high contrast pixels exist underneath an inferred line. However, this method has several problems. One issue is if characters in a cell are spaced out significantly, a vertical inferred line could exist in between two characters and the cell's contents would be split into two cells.

While these concerns can be partially alleviated by looking a certain distance to the left and right for high contrast pixels, this introduces a new risk of merging cells unnecessarily. Additionally, merging with that method can be triggered by noise in the tables. For these reasons, a neural network has been created to handle the possible merging of cells.

The CNN takes in two adjacent cells as inputs and resizes them to be $100 \times 100$ pixels each. Testing shows that this size allows the OCR to accurately extract data. The cells are resized independently so that the line that divides them is always in the center of the two cells. If instead the cells are merged and then resized, the CNN would need to predict which data belongs to which cell or it would require the location of the dividing line as an additional input. These approaches result in significantly lower accuracy and higher complexity, respectively.

The proposed CNN outputs the chance that these two cells are merged and if one or both of the cells are empty. By using a neural network to compute this information, the effect of imperfections in the image can be greatly reduced. It is important to know which cells are empty to avoid running OCR on those cells. This increases performance, as the OCR is the most time consuming step. The following sections talk about the implementation details of the CNN, the training data, and the post CNN concatenation.

\begin{figure}
\begin{center}
\includegraphics[width=.48\textwidth]{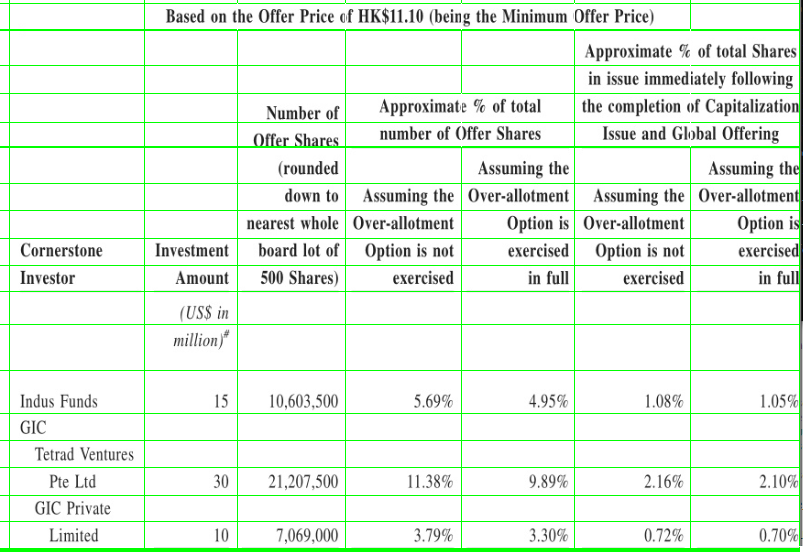}
\caption{(Table B) The table after the CNN merged problematic cells. Rows kept are separated for accurate OCR.}
\label{fig:extract_final_fixed}
\end{center}
\end{figure}
\begin{figure*}
\begin{center}
\includegraphics[width=0.8\textwidth]{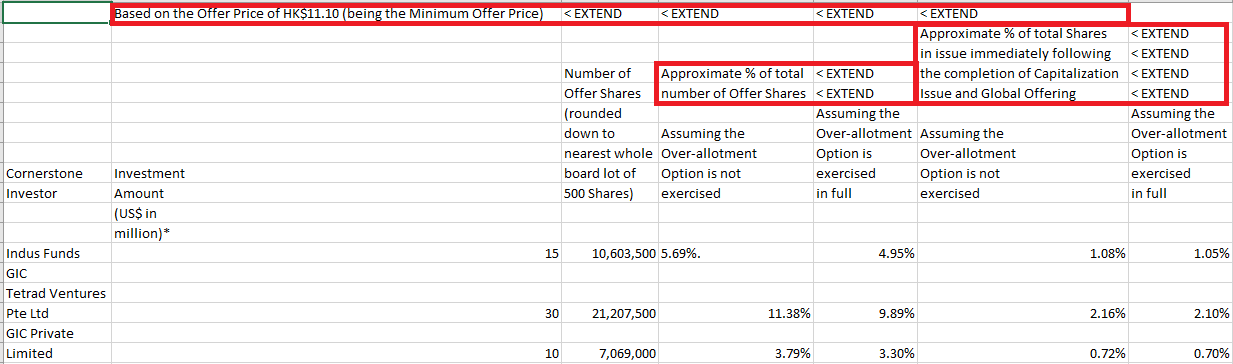}
\caption{(Table B) Tablext output of the table in Figure~\ref{fig:extract_original}, in CSV format. Thick boxes (red) show the horizontally merged cells. Cells with multiple rows are kept separated for readability.}
\label{fig:extract_csv}
\end{center}
\end{figure*}

\subsubsection{CNN Construction}
The two cells that have been sent into the CNN are merged and rotated so that the potential improper line separating these two is always a vertical line in the center of the merged image. The merged image, as the CNN sees it, always has a dimension of $200 \times 100$.

The CNN has three branches that are concatenated into a dense network. The first branch has three 2D convolution layers with $16$ filters with a size of $3 \times 3$, each layer is directly followed by a maxpooling layer of size $2 \times 2$. This branch's main purpose is to give low fidelity information around the possibly improper line that splits the cells, but it also helps identify which cells contain data.

The second branch begins with an asymmetric averagepooling layer of size $1 \times 100$. The output of this layer has the dimension of $200 \times 1$. This layer is followed by five 1D convolution layers with four filters each and a size of $3$. The averagepooling gives a good noise resistant indication if any data exists within the particular columns. This information helps identify the existence of data within one or both of the cells, but also helps confirm the presence of an improper split of the two cells.

The third and final branch does not take in the full merged image like the first two, instead it takes in the centermost $20 \times 100$ pixels. The exclusive purpose of this branch is to identify an improper splitting of the cells. Intuitively, the pixels where the cells are merged and the surrounding pixels contain the most crucial information regarding whether the cells should be merged. More complex neural network layers are used in this branch due to the importance of the information and the limited amount of inputs. Five convolutions layers, each with $64$ filters of size $3 \times 3$, are stacked upon one another.

All of these branches are individually passed through a dropout layer with a $20\%$ dropout rate before they are flattened and concatenated. The concatenated data is sent into a dense layer of size $256$, which is followed by another dropout layer, with a dropout rate of $50\%$ and another dense layer of equal size. This second dense layer is then connected to the final output which contains $3$ nodes. The first two nodes signify if the two cells contain data and the third node is high when two cells should be merged. The first dropout layers help guide the network away from relying purely on a single branch. Additionally, all of the dropout layers help avoid overfitting that would otherwise be present due to the limited amount of training data.

\subsubsection{CNN Training}
To train this CNN, a dataset that includes over $1000$ tables from various domains is procured and manually annotated. The groundtruth of every table's location, along with its respective cells' locations, is stored within Extensible Markup Language (XML) files. To translate this dataset into useful data for training purposes, Tablext's line identification techniques are used to split up the cells. All possible pairs of adjacent cells are then merged and these merged cells are used as the training data. 

To create correct labels for training, the cells' locations are used. If an active cell, according to the XML data, exists on either half of the merged cell, the respective half is considered to have data. When the XML data signifies that a single cell covers some part of both halves of the merged cell, the label for concatenation is high.

\subsubsection{Post CNN Concatenation}
The outputs of the CNN fill in matrices that record which cells should be merged and which cells contain data. Multiple cells in both the vertical and horizontal directions can be combined, if necessary, into a single large merged cell. Figure~\ref{fig:extract_final_fixed} is a debug image that shows this result. Cells with multiple independent horizontal lines are not combined at this stage, so that the OCR is fed a single line of data for high accuracy. These horizontal lines can be merged later to allow for a more concise table, or left separated to better represent a table's proportions. With the cells properly defined, the data is almost ready for OCR. However, the resolution used by the table extractor for efficiency, a width of $800$ pixels, is generally not sufficient to obtain accurate results from OCR. To address this problem, the lines defining the cell boundaries are scaled and used to cut cells into the original image.

\subsection{OCR}
All the non-empty, individual cells are sent to Tesseract, an open-source OCR tool. The returned text is then stored in a Column Separated Values (CSV) format with the location provided by Tablext. The output is shown in Figure~\ref{fig:extract_csv}. All the data within an extended cell is placed within a single cell in the CSV. To allow for accurate data interpretation, all of the merged cells point back to that original data by using the keyword ``EXTEND" and a directional arrow. These cells are emphasized with boxes in Figure~\ref{fig:extract_csv}. For clarity, cells that take up multiple horizontal lines are not merged in the CSV.

\section{Evaluation}
\label{sec:Results}
Most state-of-the art table extraction papers do not have their code available for download. However, both DeepDeSRT and TableNet tested their design on the ICDAR 2013 table competition dataset \cite{ICDAR2013}. To compare Tablext quantitatively to these previous works, Tablext is also tested against this dataset. Additionally, Tablext's extraction capability is compared against the popular, open-source, PDF extractor Tabula~\cite{tabula} on a manually procured diverse dataset with over $100$ tables. Although table identification is not the focus of this paper, the proposed method produces high quality results, while only using a small fraction of the total runtime.

\subsection{Table identification results}
In order to test the performance of the table identifier, a custom dataset is procured with over $400$ tables from various domains. This dataset includes many abstract tables without any lines that are much harder to identify than most tables.

Let AP be the area predicted in an image and AL be the true occupied area. Precision and recall, for an image, can be calculated by $\textrm{precision} = \frac{AP \cap AL}{AP}$ and $\textrm{recall} = \frac{AP \cap AL}{AL}$.
The results of the custom CNN, the YOLO model, and a combination of the custom CNN and YOLO model have been summarized in table \ref{table:yolo_comp}. 

Although the YOLO only model achieves higher precision, the higher recall of the combination of the custom CNN and YOLO is preferable since it is better to over propose regions early on than to miss any tables. This is because regions that are not tables can be discarded later in the extraction process while completely missed tables cannot be recovered. 

\subsection{Table extraction result comparison}
\subsubsection{ICDAR comparision}
Both DeepDeSRT and TableNet use $34$ randomly selected images from the ICDAR dataset for their testing set. They then calculate precision and recall in the specific way mandated by ICDAR. For every cell within a table, ``adjacency relations", defined by ICDAR, are found with its nearest horizontal and vertical neighbors. In order to get precision and recall, the adjacency relations are compared with the ground truth. Precision and recall are computed for each document then the average is taken across all the documents. 

To compare Tablext to the other papers, the same method to calculate precision and recall is used. It should be noted that both DeepDeSRT and TableNet fine tune their networks by training on the remaining ICDAR dataset. In order to prove its ability to extract data from tables in general formats, Tablext does not use any ICDAR data to train with. The results comparing the different methods can be found in Table~\ref{table:ICDAR_Results}. Tablext has both the highest precision and overall F1-Score out of the methods.

\begin{figure}
\begin{center}
\includegraphics[width=\columnwidth]{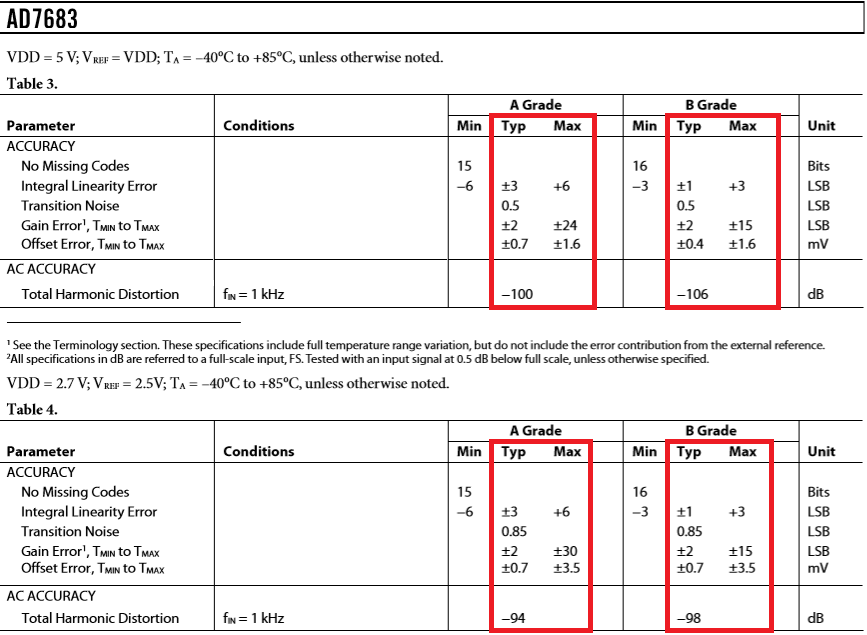}
\caption{(Table D) A text based PDF page in our testbench with potentially problematic columns highlighted with boxes (red).}
\label{fig:Tablext_PDFtest}
\end{center}
\end{figure}

\begin {table}[t]
\caption{Table Identification YOLO Comparison.}
\centering
\label{table:yolo_comp}
\begin{tabular}{|c||c|c||c|}
\hline\hline
            &  Recall & Precision & F1-Score \\
\hline\hline      
        Custom CNN Only & $0.5093$  & $0.3802$  & $0.4354$  \\
\hline
       YOLO Only &  $0.8654$  &  $0.8989$  & $0.8818$  \\
\hline
        Custom CNN + YOLO & $0.8716$ & $0.8663$ &$0.8689$  \\
\hline

\end{tabular}
\end {table}

\begin {table}[t]
\begin{center}
\begin{threeparttable}
\caption{ICDAR Data Extraction Results Comparison.}
\centering
\label{table:ICDAR_Results}
\begin{tabular}{|c||c|c||c|}
\hline\hline
        & Recall & Precision  & F1-Score \\
\hline\hline       
    Tablext & $0.9091$ & $0.9221$ & $0.9156$ \\
\hline
    DeepDeSRT\tnote{*} & $0.8736$ & $0.9593$ & $0.9144$ \\
\hline
    TableNet\tnote{*} & $0.9001$ & $0.9307$ &  $0.9151$ \\
\hline
\end{tabular}
\begin{tablenotes}\footnotesize
\item[*] Trained on a subset of the ICDAR dataset.
\end{tablenotes}
\end{threeparttable}
\end{center}
\end {table}

\begin {table}[t]
\caption{Diverse Data Extraction Results Comparison.}
\centering
\label{table:PDF_result}
\begin{tabular}{|c||c|c||c|}
\hline\hline
            &  Recall & Precision & F1-Score \\
\hline\hline      
        Tablext  & $0.9192$  & $0.9437$  & $0.9313$ \\
\hline
        Tabula   &  $0.7110$  &  $0.734$  & $0.7223$   \\
\hline
\end{tabular}
\end {table}

\subsubsection{Diverse Dataset comparison}
\begin{figure}
\begin{center}
\includegraphics[width=.4\textwidth]{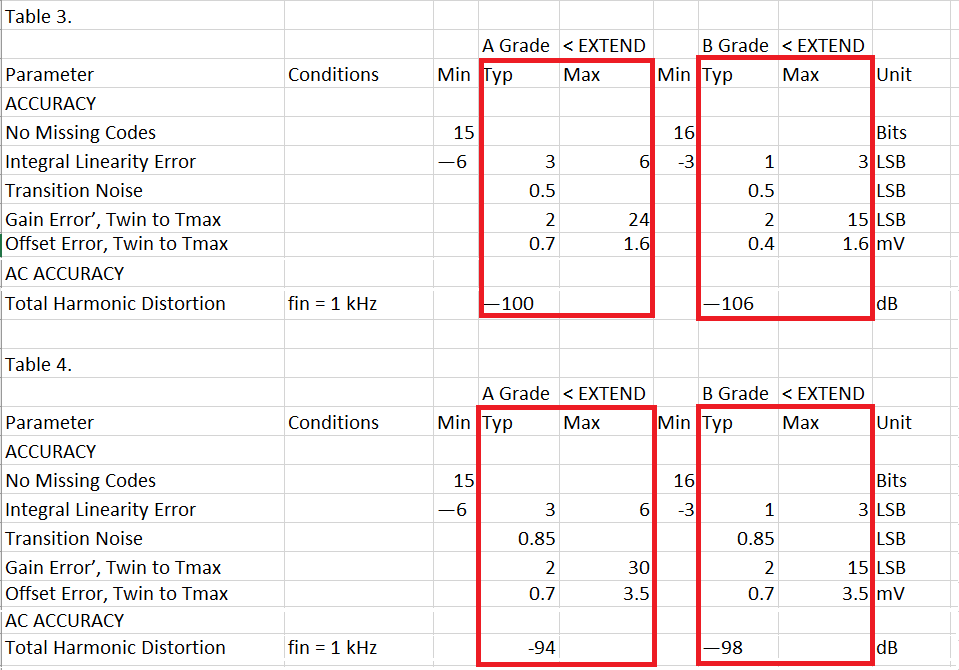}
\caption{(Table D) Tablext's final output in CSV format. Boxes (red) are showing the correct handling of problematic cells.}
\label{fig:Tablext_PDFtest_csv}
\end{center}
\end{figure}

\begin{figure}
\begin{center}
\includegraphics[width=.4\textwidth]{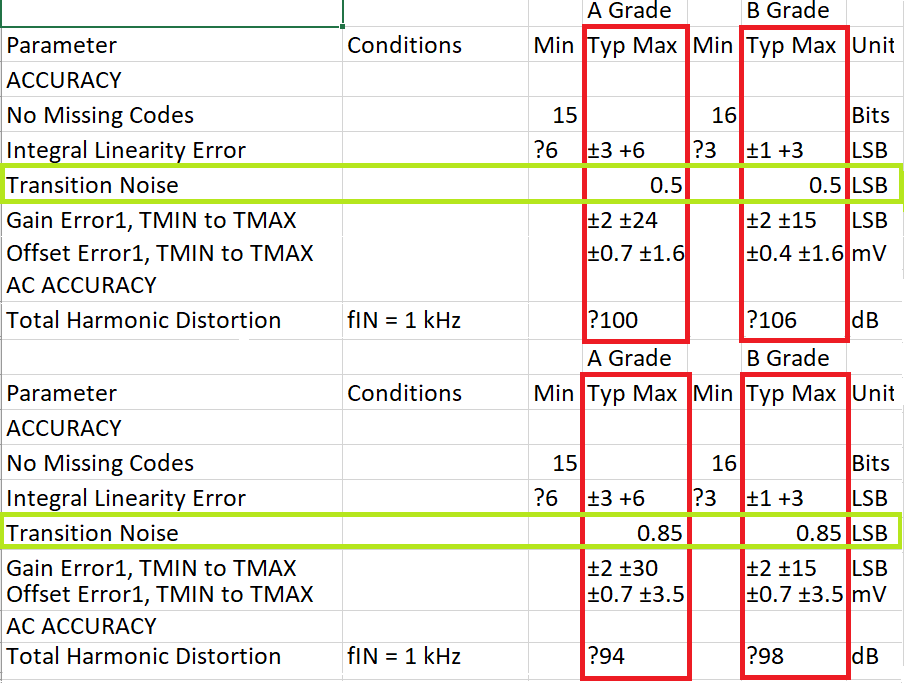}
\caption{(Table D) Tabula's final output in CSV format. Boxes (red) are showing the incorrect handling of problematic cells}
\label{fig:Tabula_PDFtest_csv}
\end{center}
\end{figure}

Without the source code for DeepDeSRT or TableNet, Tablext cannot compare to them qualitatively or quantitatively on a large diverse dataset. Therefore, a comparison is made with the popular PDF extractor Tabula. Tabula can only extract non-scanned, text-based PDFs. So a testbench, that exclusively contains text-based PDFs, is created by randomly selecting over $100$ tables within PDFs from various domains. Tabula is given the PDF pages with the text and meta information, while Tablext is given a scanned image of each PDF page. Figure~\ref{fig:Tablext_PDFtest} is an example of one such page within our testbench.

Figure~\ref{fig:Tablext_PDFtest} is extracted by both Tablext and Tabula. With this input, Figure~\ref{fig:Tablext_PDFtest_csv} and Figure~\ref{fig:Tabula_PDFtest_csv} respectively show Tablext's and Tabula's final CSV output. Both have high accuracy when reproducing the cells' contents, though Tabula is simply reading text provided to it, while Tablext is using OCR. One issue with the Tablext output is that Tesseract appears to not recognize the plus-minus character. Tabula also did not perfectly print the data within the cells, despite reading straight from the PDF. The minus character could not be recognized by Tabula, this problem appears because this PDF is using an abnormal character instead of the standard ASCII symbol.

While both methods have high accuracy when reproducing the cells' contents, the same is not true about them placing the content. Several severe structural errors appear in Tabula's output and these errors have been highlighted in Figure~\ref{fig:Tabula_PDFtest_csv}. Looking back at the original image in Figure~\ref{fig:Tablext_PDFtest}, it becomes clear what caused these errors. The cells ``A Grade" and ``B Grade" overlap both the columns beginning with ``Typ" and ``Max".  Errors like this are common for conventional text based extractors~\cite{dreslinski2019fully}. Tablext meanwhile, with its dynamic threshold ratio, easily separates these two columns then merges the split ``A Grade" and ``B Grade" cells back into a single cell.

The highlighted cells in Figure~\ref{fig:Tabula_PDFtest_csv} emphasize the importance of accurate cell location identification. Even with clever post-processing, it would be impossible to tell which column the merged cells with a single piece of data belong to. For instance, looking at the row beginning with ``Transition noise", there is no way to know if the data belongs to the ``Typ" or ``Max" column.

The results comparing the two methods can be found in Table \ref{table:PDF_result}. Tablext's ability to handle complex tables greatly surpasses Tabula's, despite all of the extra meta information that Tabula has access to.

\section{Conclusion}
\label{sec:Conclusion}
This paper introduces a novel, general approach for table extraction. By utilizing both deep learning and computer vision techniques, the neural network can focus its attention on complex problems, while allowing more conventional methods to handle the simpler tasks. This focus allows Tablext to beat competing state-of-the-art neural network based table extraction methods as well as a popular open-source tool that requires table meta data.

\section*{Acknowledgment}
This material is based on research sponsored by Air Force Research Laboratory (AFRL) and Defense Advanced Research Projects Agency (DARPA) under agreement number FA8650-18-2-7844. The U.S. Government is authorized to reproduce and distribute reprints for Governmental purposes notwithstanding any copyright notation thereon.

\bibliographystyle{IEEEtran}
\balance
\bibliography{Bibfile}
\end{document}